# System Modeling in the COSMA Environment

W. B. Daszczuk, W. Grabski, J. Mieścicki, J. Wytrębowicz

Institute of Computer Science, Warsaw University of Technology
ul. Nowowiejska 15/19, Warsaw 00-665, Poland,

**Abstract**
*The aim of this paper is to demonstrate how the COSMA environment can be used for system modeling. This environment is a set of tools based on Concurrent State Machines paradigm and is developed in the Institute of Computer Science at the Warsaw University of Technology. Our demonstration example is a distributed brake control system dedicated for a railway transport. The paper shortly introduces COSMA. Next it shows how the example model can be validated by our temporal logic analyzer.*

## 1. Introduction

Modeling is an obligatory step during design of any system, which is crucial for people safety or whose unreliability can lead to important lost of money. This step is recommended during design of any large size system, where it is the only way to define a complete functionality of the system. A designer performs the behavioral modeling of software implemented, hardware implemented or mixed systems. The implementation details have no regard at this step. Redundancies of an implementation increase its reliability, however they cannot resolve conceptual deficiencies. The first important labor is to build a complete behavioral specification that is correct with respect to demands of the system users and its environment. We call this step validation of the system.

Unfortunately we can notice, that in practice the modeling is frequently neglected. One of the reasons is the lack of inexpensive and easy to use modeling platforms for designers. This is the motivation to develop in our Institute such a platform that we call COSMA. The principle of Concurrent State Machines is the basis for modeling of reactive, discrete and control oriented systems. Due to this paradigm it is possible to build analytic tools, which can process designer questions. In COSMA, the designer formulates them in a TLA notation. The responses approve the desired model behavior or show its incompleteness, overhead or conceptual errors.

There is other environment that allows to perform similar tasks. It is the Promela specification language and the Spin validation engine developed by Gerard Holzmann at Bell Labs [18] [5]. That environment is older and more matured than COSMA. However our intention in the COSMA development was to make it more general. As Spin/Promela is well adapted for software system validation, especially for communication protocols, the COSMA environment is appropriate for modeling hardware [7] and software [6] as well.

Many formal verification methods has been proposed and vast literature describes them [4] [8] [9] [10] [16]. However, still small number verification environments exists, and more less are used by engineers. This paper is not devoted to compare them. It reports just a case study in the aim to introduce a new tool, which is developed in pragmatic view, using modern algorithms to enable an elaboration of a huge state space model.

## 2. COSMA tools

We develop an original software toolset COSMA 2.0 in the Institute of Computer Science (Warsaw University of Technology). The main part of its present version consists of three modules: Grapher, Product Engine and TempoRG.

The conception of Concurrent State Machines (CSM) [3] [13] is the basis for COSMA environment. The Concurrent State Machines (or automata) are labeled, directed graphs, which can be abstract models of discrete objects, e.g., control units, programs, processes, protocols etc. The ultimate goal of this modeling is the analysis or verification of the behavior of a system of cooperating, concurrent components.

A designer assigns one automaton for every structural sub-unit of his system, as well as the communication connections among them. Then, the CSM models of individual system components have to be developed.

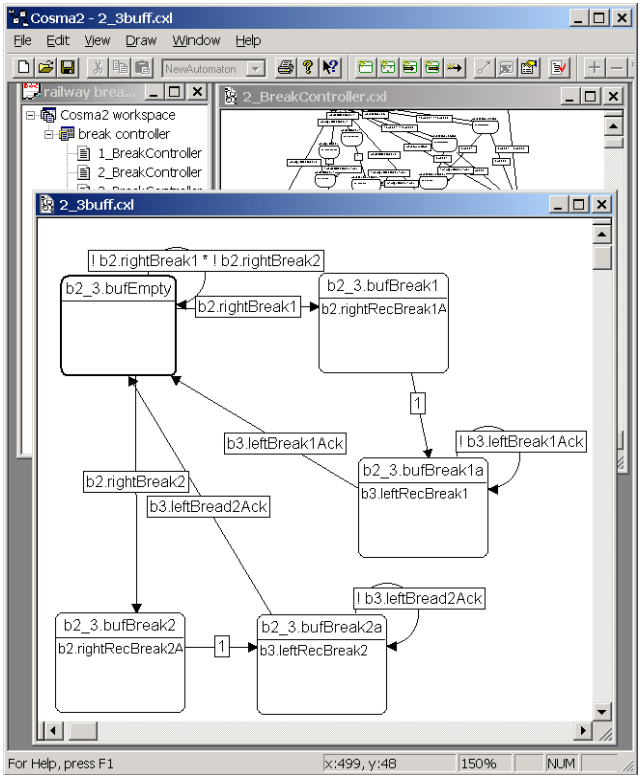

Figure 1. Grapher window at COSMA environment.

The Grapher module provides the user interface for drawing and editing CSM models. It also converts graphical specification of system components into XML-like language called CXL [15]. Figure 1 shows a screen snap from Grapher.

At a first glance, a module resembles a typical and well-known Moore finite automaton. However, in contrast to conventional FSM, in CSM arcs are labeled with Boolean formulas rather than with symbols from an input alphabet. For instance, formula a would mean that 'symbol a occurs at machine's input', !a*!b means that 'neither a nor b occurs' etc.

The arc (s, s') from node s=*b2_3bufEmpty* to s'=*b2_3bufBreak1*, labeled with formula f=*b2.rightBreak1*, means that s' can follow s if (and only if) formula f is true. If s=s'=*b2_3bufEmpty* (i.e., an arc makes a 'loop' over the same state), then formula f=! *b2.rightBreak1* * ! *b2.rightBreak2* represents a condition under which the machine can remain in s. Otherwise (i.e., if s ≠ s)', the arc represents a transition (from s to s') while its formula f specifies a condition that enables this transition. Note that two or more Boolean formulas can be simultaneously true and – consecutively - more than one arc from a state can be simultaneously enabled. Then, only one of them is selected. The choice is non-deterministic. Note also that arcs labeled with the condition 1 ('unconditionally true', by the definition) can be used. They are interpreted as spontaneous transitions that require no external events or messages to be enabled.

Thus, Concurrent State Machines represent the conditions for changes of states in terms of occurrences of abstract symbols from some finite input alphabet. The practical interpretation depends on the nature of a system under consideration. In a model of communicating software processes, 'symbols' may stand for specific events, messages or conditions. In hardware models, symbols are usually interpreted in terms of logical values assumed by binary variables. For example, the formula 'ready*!bbsy' would mean that the transition has to be executed 'if bus line ready is set to 1 while bbsy is reset to 0'. The use of abstract symbols instead of application-specific conventions is an advantage of the CSM model, as it provides the common framework for the specification of both hardware or software structures.

This way of symbol perceiving differs our CSM model from other formal specification techniques such as Estelle, SDL or LOTOS [16], where the symbol interpretation is fixed. The automata communicate via messages. Thus the hardware signals cannot be modeled in a simple manner. A message can be treated as signal transition, but this overcome make the specification bigger and more difficult to understand.

The key point in the CSM model is that (again in contrast to conventional FSM) the sequential occurrence of input symbols is not assumed. Input symbols are not 'pre-synchronized' (e.g., sequenced or interleaved) in any way. At any instant of time, they can come either alone or simultaneously or even not come at all. Moreover, any component of a system can transmit its own output symbol that can be an input to neighboring machines (and even to itself). No implicit synchronization among component's activities is assumed. This way, the CSM model supports communication among mutually asynchronous, concurrent system components and their environment.

A computation of so-called Reachability Graph (RG) of the system is the key element of the CSM modeling. RG is a tree of all system states reachable from the system's initial state along with appropriate Boolean formulas.

The Product Engine module converts the CXL specification into a set of Binary Decision Diagrams (BDD) and then computes the system's Reachability Graph, which is again a large BDD. This module uses the state-of-the-art library of functions for processing ROBDDs, implemented by Geert Janssen from Eindhoven University of Technology.

The Product Engine module computes the product of individual models of components. In that way it generates all configurations or co-incidences of component states and all transitions that are likely to occur. This product constitutes input data for the next COSMA tool:

the TempoRG module that contains a set of algorithms for the evaluation of temporal requirements in a given RG of a system.

The analysis of RG may detect and identify harmful synchronization errors, like a deadlock, a livelock, possible lack of response for some specific event, unwanted simultaneous activity of two components etc. These errors (practically unavoidable in the design of asynchronous and concurrent structures) are hardly detectable by simulation and testing, as they may result from very rare coincidences of components' states and external stimuli. RG includes all practically possible states and transitions, therefore it highlights even very rare sequences of events.

In general, RG can be of an enormous size, which causes well-known time and space complexity problems. In the case of simple systems analyzed just for tutorial purposes one can draw or print the RG and analyze it 'by hand'. In more practical cases, the number of RG nodes (i.e., system states) can be of order of $10^{20} - 10^{50}$ or even more [2]. To manage the problem, large graphs are usually represented in a form of data structures known as ROBDD (Reduced Ordered Binary Decision Diagrams [1]) that allow for very concise representation. Due to this, in many practical cases the development and analysis of system's RG does not exceed storage and processing power capabilities of an average workstation.

The inspection of such a large RG cannot be done 'by naked eye'. Thus, one should formally specify the requirements for system's behavior and then use the appropriate algorithm for the evaluation if these requirements are actually satisfied in a given RG. The commonplace approach involves the use of temporal logic, where the requirements have the form of temporal formulas. There are many types of temporal logic, but generally they allow for constructing sentences, where temporal connectives (always, eventually, next, until) can be used in addition to 'classical' Boolean operators (not, and, or, if ... then ... etc.) and two quantificators (for all ..., exists ...). Temporal propositions expressed this way can cover a very wide class of requirements addressing the issues of the flow of control, communication and synchronization among components.

COSMA is already a powerful symbolic model checker. The Grapher allows to draw hierarchical states. A hierarchical state is a shortcut of an automaton piece. Using hierarchical states, the designer can hide some specification details or visualize in more readable form his system. We still enhance COSMA functionality adding new modules, currently among them are:
- Translator from UML state diagrams to CSM,
- Extended CSM (ECSM) grapher and simulator.

ECSM enhances the expressive power of CSM, as it allows for specification of general data structures and arbitrary operations on these data. Operations on data can be attributed either to states or to transitions of the CSM, which becomes this way just a scheme of flow of control in a process. ECSM models are analyzed by simulation. This way, in the design process of a concurrent, asynchronous system one can verify the correctness of communication and synchronization mechanisms by finite state model checking and evaluate the system's performance by simulation of its ECSM model as well.

The future enhancements planed for the third COSMA version consists of:
- Behavioral Constraint Language front-end to the TempoRG module,
- generators of implementation skeletons in C programming language and Verilog hardware description language.

We plan to add new algorithms to the TempoRG analyzer. The Behavioral Constraint Language will simplify their use. System modeling is not an end of design process. Next the designer starts to build its implementation. Having the analyzed model in mind, he builds the implementation faster and without conceptual errors. Although the passing from CSM model to an implementation skeleton is not a heavy task, the planed generators could speedup this step.

## 3. Model of a distributed brake control system

Our case study is a simplified, distributed brake control system for railway transport. The system consists of independent controllers that communicate to each other. Every car of a train has one controller, which activates brakes of the car and selects a brake force. The controller obtains signals from:
- velocity meter,
- coils detecting distance from a station (there are activators placed longwise the rails),
- emergency brake levers.

It obtains also messages form one or two controllers, which are situated in the next and previous car of the train. The messages indicate the selected activity for local brakes of the neighbor cars. While an indication of higher brake force arrives than actually selected, in that case the higher force is applied for local brakes.

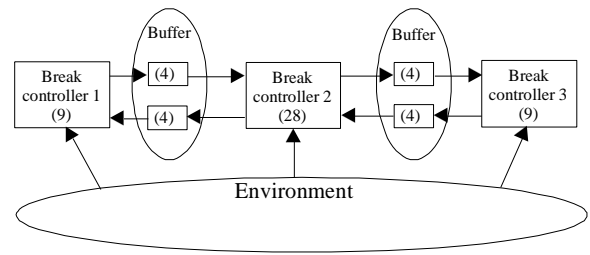

Figure 2. Model structure of the distributed brake control system.

The structure of the system is straightforward, we assign one automaton for every controller, Figure 2. The controllers communicate via messages. To model a communication medium with associated message buffers, we define an automaton. A simple modification of this automaton will be sufficient for unreliable medium modeling. Because signals from the velocity meter and from distance detectors are not hazardous, we need simple automata to model their behavior, especially to express their possible malfunctions. However in the early modeling steps we can assume that they belong to the system environment and the signals they generate are external. This assumption gives a smaller state space of the generated reachability graph. Figure 2 shows numbers in parenthesis, which are state numbers of every automaton. The multiplication of them all gives 580 608 states. Our Product Engine generates 9 061 reachable states.

A complete brake control system can assemble any number of controllers. The behavior of three, four or more controllers is the same. To minimize the size of total state space only three controllers we take into consideration. Of course the simpler cases, i.e., one and two controllers in the system, we should model too.

We distinguish principal states of the controller, which are related to the applied break force. Similarly to the number of controllers, we do not have to model a big number of different break forces - three is enough to verify the concept of our system. They are:
- break0 - break force is 0,
- break1 - middle force,
- break2 - big force.

| b2_commCord | Emergency brake lever pulled |
| --- | --- |
| b2_leftBreak2 | Message "force 2" to the left controller |
| b2_rightBreak2 | Message "force 2" to the right controller |
| b2_leftRecBreak2Ack | Acknowledgement form the left transmission buffer |
| b2_rightRecBreak2Ack | Acknowledgement form the right transmission buffer |
| b2_leftRecBreak2 | Message "force 2" from the left controller |
| b2_rightBreak2Ack | Message "force 2" from the right controller |
| b2_distSens1 | Car passes the $1^{st}$ distance activator |
| b2_distSens2 | Car passes the $2^{nd}$ distance activator |
| b2_distSens3 | Car passes the $3^{rd}$ distance activator |
| speed0 | Current car speed is low |
| speed1 | Current car speed is medium |
| speed2 | Current car speed is high |

*B2_commCord* reception (emergency brake) causes the

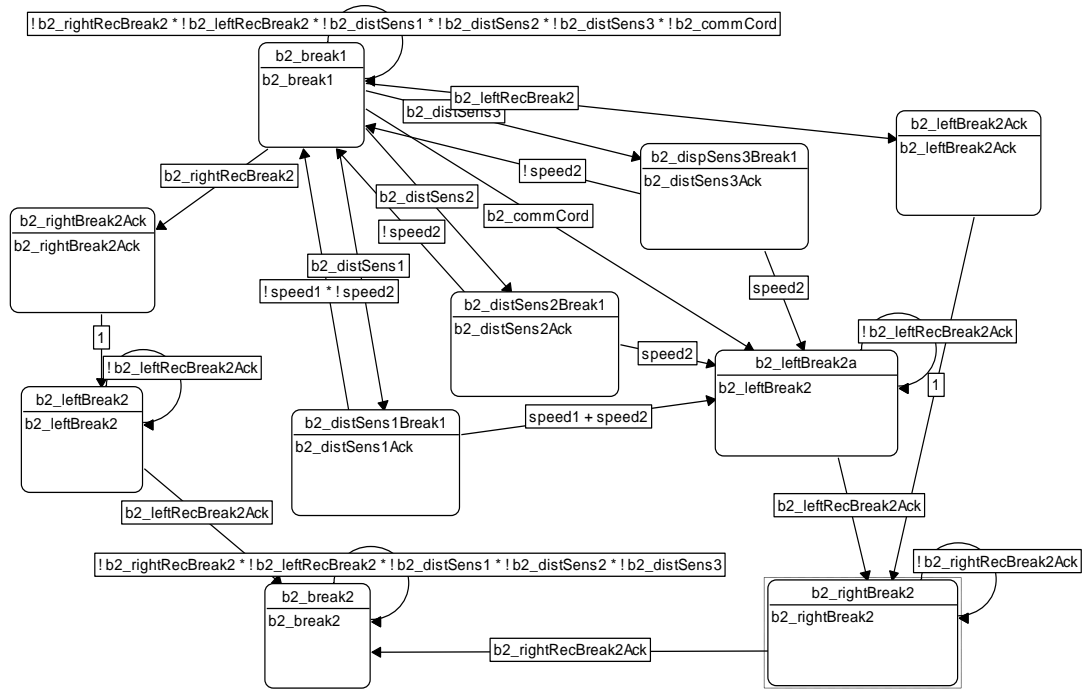

Figure 3. Part of the break controller model.

Other states are related to signal and message reception. Figure 3 depicts a part of this automaton. It shows two principal states (i.e., *b2_break1, b2_break2*) and eight transitional states, which control signal exchange.
Table I explains the meaning of all depicted signals.
TABLE I - Interpretation of signal from Figure 3

| SIGNAL | MEANING (when *true*) |
| --- | --- |

transition form *b2_break1* to *b2_break2* through the states: *b2_leftBreak2a* and *b2_rightBreak2*. At these states, messages to the neighbor controllers are generated and acknowledges are awaited.

*B2_leftRecBreak2* reception from the left neighbor (his brake force is 2) causes the transition form *b2_break1* to *b2_break2* through the states: *b2_leftBreak2Ack* and

*b2_rightBreak2*, where respectively an acknowledgement to the left and message (force is 2) to the right are generated.

In the same way is processed *b2_rightRecBreak2* signal from the right. In this case the intermediate states are: *b2_rightBreak2Ack* and *b2_leftBreak2*.

Distance detector generates three signals: *b2_distSens1, b2_distSens2, b2_distSens3*. Reception of one leads to the respective state *b2_distSens1Break1*, *b2_distSens1Break2* or *b2_distSens1Break3*, where an acknowledgement is generated. The following transition depends on the actual train speed and can conduct to the states:
*b2_break1* - no change of brake force,
*b2_break2* - through the states *b2_leftBreak2a* and *b2_rightBreak2*, where respective messages to the neighbors are generated.

We have defined the rest of the model, not depicted on Figure 3, in similar fashion as described above. The Product Engine generates a reachability tree, and we can display it in a COSMA window. However due to the tree size it is better to analyze it by our TempoRG analyzer than by a naked eye.

## 4. Validation of the brake control system

The first almost ritual question asked in every validation or verification process is: "Is there any possible deadlock in my system?" This question applied to the CSM model can be formulated: "Is there any state, when no other state can be reached?" We write it in the temporal logic notation:
   ?- s: ◊ (¬ in s)
The analyzer answer is:
   --> FULFILLED FOR STATES:
       ALL
   Evaluation time is 00:00:15/422 (15422 ms)
This answer means that there is no deadlock in our specification. The next question every designer likely ask is about a livelock. In the CSM model this is the question about cycles: "Find strongly connected subgraphs." We write it in the temporal logic notation:
   ? s: ○ ◊s
There are about 500 cycles, listed by analyzer:
   --> FULFILLED FOR STATES:
   b1_2_bufEmpty:b1_break0:b2_1_bufEmpty:b2_3_bufEmpty:b2_break0:b3_2_bufEmpty:b3_break0
   Evaluation time is 00:06:34/207 (394207 ms)
   …
   b1_2_bufBreak1a:b1_distSensBreak2Right:b2_1_bufBreak1a:b2_3_bufEmpty:b2_distSensBreak1Left:b3_2_bufBreak1a:b3_break0
   Evaluation time is 00:00:00/00 (0 ms)
The analyzer cannot distinguish itself, which cycle is a correct and which one is a lifelock. Designer should resolve this problem by post processing the obtained cycles or by asking a more precise question. In our example case we have not found any livelock.

Next designer asks questions about system functionality. More question will be evaluated, better, more complete verification will be done. An example question we have asked was: "Do all cars will brake with force 2, if someone pull emergency brake lever?" The question in temporal logic is:
   G (in AUT_5.b2_distSensBreak2Left => (◊ (in AUT_2.b1_break2 ∧ in AUT_5.b2_break2 ∧ in AUT_7.b3_break2 ) ) )
The analyzer answer conforms our expectations:
   --> TRUE
   Evaluation time is 00:53:58/667 (3238667 ms)
Other question we have asked was: "Does the train will start moving after a break?" or in a simpler form "Will the middle controller apply force 0 to its break after applying force 2?"
   G ((in aut_5.b2_break2) => (◊(in aut_5.b2_break0 )))
Unfortunately the answer is:
   --> FALSE
   Evaluation time is 00:02:49/234 (169234 ms)
We have found an error in our model - the lack of transition from state "break force 2" to the state "break force 0" when the train speed is equal 0. After correction of this error the analyzer have returned the expected answer.

## 5. Conclusions

The CSM model and the methodology based upon the COSMA environment have been developed for system level design of asynchronous, cooperating circuits.
The paper not only introduces the new modeling environment, but also gives a tutorial view on formal verification process. Summarizing our experiences, we would like to emphasize that:
- Specification of the behavior in terms of the CSM model is easily understandable and close to the common intuition.
- CSM model is a formal one and supports the formal verification of system's behavior.
- The CSM specification facilitates the generation of ECSM model that can be simulated, e.g., to analyze performance parameters.
- The CSM specification facilitates the implementation process owing to artful understanding of the system behavior and environment constrains.

We plan two directions for future COSMA evolution. The first is to make TempoRG algorithms more powerful and the human interface for this analyzer more user friendly. The second direction is to make a link to the popular CASE tools, with the aim of helping designers to build correct systems.